\documentclass[12pt,preprint,pra,showpacs,showkeys,amsmath,amssymb]{revtex4-1}
\usepackage{graphicx}
\usepackage{dcolumn}
\usepackage{bm}
\usepackage{epsfig}
\usepackage{color}

\begin{document}
\title{Effect of pore-size disorder on the electronic properties of semiconducting graphene nanomeshes} 
    
\author{Sarah Gamal$^{1}$}
\author{Mohamed M. Fadlallah$^{2,3}$}
\email{mohamed.fadlallah@fsc.bu.edu.eg} 
\author{Lobna M. Salah$^{1}$}
\author{Ahmed A.\ Maarouf$^{4}$}  
\email{amaarouf@iau.edu.sa}  
\affiliation{
$^1$Department of Physics, Faculty of Science, Cairo University, Giza 12613, Egypt}
\affiliation{
$^2$Physics Department, Faculty of Science, Benha University, Benha 13518, Egypt} 
\affiliation{
$^3$Center for Computational Energy Research, Department of Applied Physics, Eindhoven University of Technology, P.O. Box 513,  5600MB Eindhoven, The Netherlands}
\affiliation{
$^4$Department of Physics, Institute for Research and Medical Consultations, Imam Abdulrahman Bin Faisal University, Dammam 31441, Saudi Arabia}

\date{\today}
\begin{abstract}
Graphene nanomeshes (GNMs) are novel materials that recently raised a lot of interest. They are fabricated by forming a lattice of pores in graphene. Depending on the pore size and pore lattice constant, GNMs can be either semimetallic or semiconducting with a gap large enough ($\sim$ 0.5 eV) to be considered for transistor applications. The fabrication process is bound to produce some structural disorder due to variations in pore sizes. Recent electronic transport measurements in GNM devices (ACS Appl. Mater. Interfaces 10, 10362, 2018) show a degradation of their bandgap in devices having pore-size disorder. It is therefore important to understand the effect of such variability on the electronic properties of semiconducting GNMs. In this work we use the density functional-based tight binding formalism to calculate the electronic properties of GNM structures with different pore sizes, pore densities, and with hydrogen and oxygen pore edge passivations. We find that structural disorder reduces the electronic gap and the carrier group velocity, which may interpret recent transport measurements in GNM devices. Furthermore the trend of the bandgap with structural disorder is not significantly affected by the change in pore edge passivation. Our results show that even with structural disorder, GNMs are still attractive from a transistor device perspective.

\end{abstract}

\maketitle

\section{Introduction}
Graphene is probably one of the most studied materials in the past decade \cite{graph1,graph2}, due to its extraordinary optical, mechanical, and electronic properties. Its high mobility (15000 cm$^{2}$/Vs) \cite{graph2,mobility} makes it a potential replacement for silicon in electronics \cite{mobility}, but a zero bandgap prevents its utilization as a field effect transistor (FET). Many approaches have been used to open a bandgap ($E_g$) in graphene based structures, such as placing a graphene bilayer in a vertical electric field \cite{electricfield}, controlled interactions with a substrate \cite{substrate}, and quantum confinement in graphene nanoribbons \cite{nanoribon}.  

Another way for opening a bandgap in the graphene spectrum is through the creation of a lattice of pores, thus forming a graphene nanomesh (GNM) \cite{mesh1,mesh2}. GNMs have been attracting research interest due to their advantages over graphene, namely their electronic and chemical properties. GNMs with specific pore symmetries and size range possess a fractional eV electronic bandgap \cite{gap}, which is very attractive to the transistor industry. Chemical functionalization is possible through controlled pore edge passivation, which can make GNMs highly and selectively reactive \cite{pass}. Pore passivation by species different from carbon creates pore edge dipoles resulting from the mismatch in electronegativity between carbon and the passivating species. On the other hand, the gap opening in GNMs is not as simple, and certain symmetry considerations have to be satisfied for a gap to open. GNMs have been synthesized with pore sizes in the range of 5-20 nm \cite{synth1,synth2,synth3,synth4}. 

Many studies have focused on the electronic \cite{electronic1,electronic2,electronic3,electronic4}, electronic transport \cite{trans1,trans2,trans3}, and thermoelectric transport \cite{hex1,hex2,heat1,heat2} properties of perfect (disorder free) GNMs. The bandgap depends on the geometrical details of the pore lattice, including the pore shape, size, and pore lattice constant  \cite{electronic1,PhysRevB.80.233405}. Generally for hexagonal pores, the bandgap is found to be proportional to the ratio between the number of removed carbon atoms at the total number of carbon atoms in the unit cell \cite{electronic1}. For example, a hexagonal lattice of 0.8 nm pores with a pore-to-pore distance of about 2.3 nm has a 0.5 eV bandgap \cite{hex1,hex2}. Highly stable {\it p}- and {\it n}-doping of GNM semiconducting structures have been predicted, where controlled pore edge passivation facilitates the trapping of electron accepting and donating species \cite{gap,hex1}. This promoted GNMs to be considered in some applications, such as FETs \cite{fet1,fet2}, molecular sensors \cite{hex1,sensor,maaroufsensorpatent}, nanoparticles support templates \cite{maaroufdockingpatent}, hydrogen storage \cite{Ali}, spintronic devices \cite{hex2}, and supercapacitors \cite{F,sup}.
 
A few techniques have been used to fabricate GNMs, including nanosphere lithography \cite{litho1}, nanoimprint lithography \cite{Liang2010}, block copolymer lithography \cite{synth3,litho3}, irradiation etching \cite{A,AA}, atomic force microscopy etching \cite{B}, and oxygen reactive ion etching \cite{C,D}. GNMs fabricated with these methods will most likely have structural defects, such as variations in the size and shape of the pores, as well as the pore lattice structure and symmetry. Structural defects can severely alter the electronic and transport properties of semiconductors. If GNMs are to be seriously considered as candidates for graphene-based transistors, it becomes crucial to understand the extent to which structural defects degrade their electronic properties, especially the bandgap size, and the carrier velocity. 

There are numerous tight binding (TB) studies of graphene based structures. The popularity of the TB method resides on its successful reproduction of the low energy spectrum of graphene and carbon nanotubes. The electronic properties of GNMs have been thoroughly studied using TB \cite{electronic1,Furst2009a,electronic2}. Disorder in the pore shape have even been considered, where it is found that for strong disorder, the bandgap varies inversely with the average pore lattice constant \cite{HungNguyen2013}, a trend that is in qualitative agreement with some experimental results \cite{Liang2010}. Despite these efforts, a DFT-based method is required to quantitatively investigate the disorder arising from various geometrical and chemical factors affecting the electronic and transport properties of GNMs. Such approach allows for a more realistic treatment of various defects arising from the fabrication process, such as local stresses and pore edge passivations, and hence provides a more accurate prediction of the bandgap and carrier velocity.

In this work, we study disorder in the pore size and pore lattice constant of hydrogen and oxygen passivated GNMs, as would be induced by the fabrication processes, using a density functional theory based TB (DFTB) approach. We study the electronic properties of GNMs with pores of multiple sizes: 0.36 nm, 0.83 nm, and 1.29 nm, and two pore lattice constants: 1.47 nm, and 2.21 nm. We calculate the density of states (DOS), band structure, and carrier group velocity. This allows us to investigate the dependence of the bandgap on various GNM parameters (pore size, lattice constant, edge passivation), as well as different structural permutations of the same pore configuration. 
 
Section II describes the methodology which is used in this work. To understand the electronic properties of {\it disordered} pores in graphene, we start in III-A with the results of GNMs having one pore size, using 3 base GNM supercells: 12$\times$12 system with 4 pores, 18$\times$18 system with 4 pores, and 18$\times$18 system with 9 pores. The results include the DOS, band structures, the dependence of the bandgap on the pore radius and the pore lattice constant. In addition, we compare the results of DFTB to two sets of corresponding results obtained by DFT, so as to validate the use of the DFTB method with our GNM structures. GNMs, with pore size and pore lattice constant disorder, and with two passivations, are discussed in III-B. In certain cases, all pore permutations of the same disordered GNM are studied. In this work, we consider a total of 56 GNM systems.

\section{Methodology}

To calculate the electronic properties of the considered large systems, we employ the self-consistent charge (SCC) density functional tight binding scheme (SCC-DFTB), as implemented in the DFTB+ package \cite{meth8}, where the total energy is a second-order expansion of the DFT energy in the charge density fluctuations \cite{meth1,meth2}. The SCC formalism describes the Coulombic interaction between atomic charges. It has been successfully used to calculate the electronic properties of bulk, surface, and nanowire systems \cite {meth3,meth4}, molecular materials \cite {meth5}, as well as some biological systems \cite{meth6}. The success of DFTB to describe the electronic properties of many systems has been demonstrated through numerous comparisons with DFT, for example to calculate the structural and electronic properties of 2D phosphorous carbide polymorphs \cite{Heller2018}, or to study the effect of structural changes on the electronic properties of GaN nanowires \cite{Ming2016}. Generally, DFTB agrees well with DFT and can therefore be utilized to study large-scale systems that are impossible to study with DFT. 

The mio Slater-Koster parameter library contains all possible atom-atom interactions in our GNM structures \cite{meth9}. The conjugate gradient algorithm is used to perform the structural relaxation of the studied GNM structures, with atomic forces less than $10^{-4}$ eV/{\AA}. Convergence with respect to the k-points grid was achieved for various GNM systems. Table~\ref{t1} shows the grid for each GNM system. A 12 {\AA} vacuum distance is used to avoid image interaction in the $z$-direction. 

\begin{table*}[htb]
\centering
{
\begin{tabular}{|c|c|c|c|}
\hline

Lattice constant& $k$-points& R (H-GNM) & R (O-GNM) \\\hline
\hline

1.47 (6$\times$6) & 60$\times$60$\times$1& 0.36, 0.83& 0.29, 0.55  \\\hline

\hline
2.21 (9$\times$9) &40$\times$40$\times$1& 0.36, 0.83, 1.21& - \\\hline

\hline
2.95  (12$\times$12) &30$\times$30 $\times$1&0.36, 0.83, 1.29& -\\\hline

\hline
3.69 (15$\times$15) &24$\times$24$\times$1&0.36, 0.83, 1.29& - \\\hline

\hline
4.42 (18$\times$18)&20$\times$20$\times$1&0.36, 0.83, 1.29& -  \\\hline

\end{tabular}
}

\caption{Lattice constant (nm), $k$-points grid needed for convergence and pore diameter, R (nm), of considered GNMs passivated by H (H-GNMs) and O (O-GNMs).} 
\label{t1}
\end{table*}

\section{Electronic structure of disordered GNM{\scriptsize s}}

There are two players when it comes to gap opening in graphene. First, there is the mixing of the distinct $K$ and $K'$ points. One can think of the pore lattice as an external potential acting on the graphene electrons. For a bandgap to open up at the Dirac points, the external potential associated with the pore lattice has to mix states at the two distinct $K$ points. For a triangular lattice of pores with a lattice parameter which is a multiple of 3 of the graphene lattice parameter, there will be a gap that increases with the pore radius. One third of these GNMs will be semiconducting \cite{gap2}. Second, there are the pore edge effects, which are akin to the bandgaps of graphene nanoribbons \cite{Son2006}. Below, we first discuss the electronic properties of H-GNMs with pores of equal size. We then move to disordered GNMs having pores of different size.

\subsection{GNMs with one pore size - disorder-free case}

\begin{figure}[h]
\includegraphics[width=0.8\textwidth]{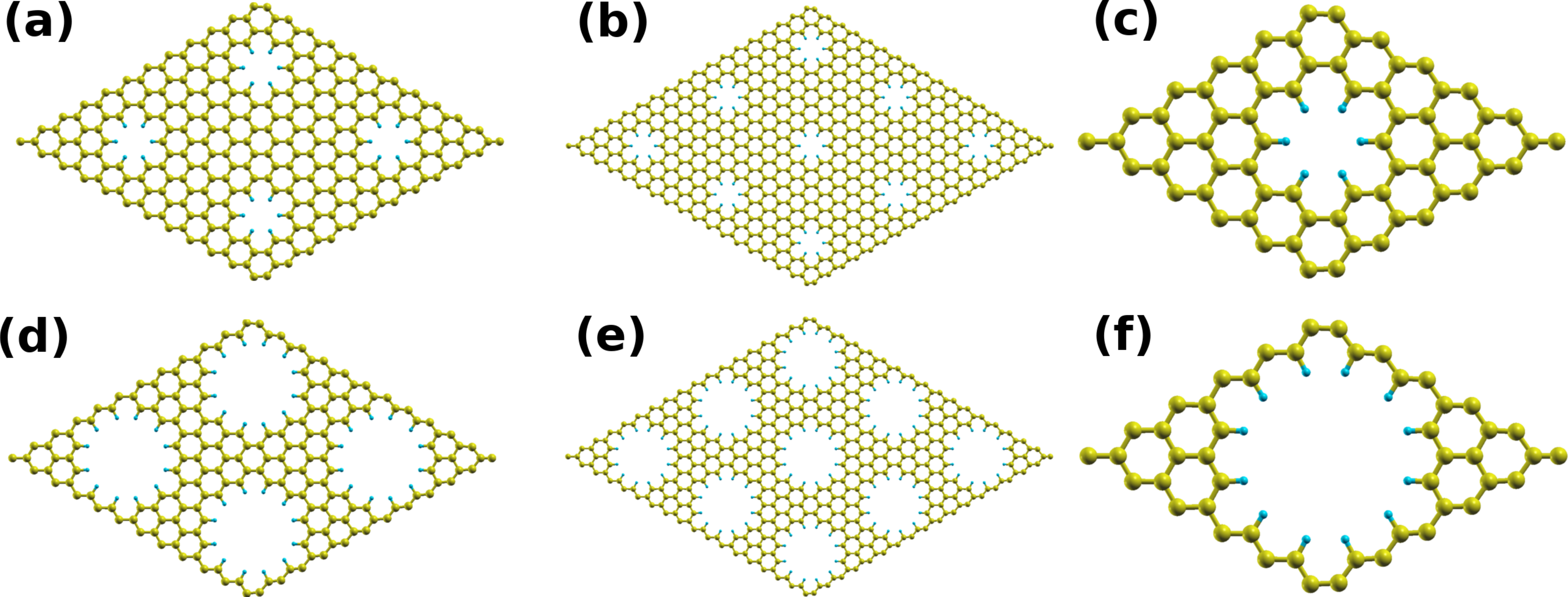}
\caption{ The H-GNMs optimized structures for (a) 12$\times$12 system made from four 6$\times$6 unit cells, (b) 18$\times$18 system made from nine 6$\times$6 unit cells, (c) 6$\times$6 system, with a small pore, (d) 12$\times$12 system made from four 6$\times$6 unit cells, (e) 18$\times$18 system made from nine 6$\times$6 unit cells, and (f) 6$\times$6 system, with a medium pore. The yellow and cyan spheres indicate C and H atoms, respectively.}
\label{fig1}
\end{figure}

We begin with H-GNMs having hydrogen passivated pores. We consider 3 pore sizes, which we term "small", "medium", and "large". Figure \ref{fig1}a shows a 12$\times$12 unit cell with 4 small pores, while figure \ref{fig1}b shows an 18$\times$18 unit cell with 9 small pores. When all pores are equal, the two systems reduce to a simpler one with the 6$\times$6 unit cell shown in Fig. \ref{fig1}c. Figures \ref{fig1}d,e and f are the corresponding systems with the medium pore. The small pore H-GNM has 8.3\% of its carbon atoms removed, while the medium pore H-GNM loses 33.3\% of its carbon atoms. Structural relaxation of the 6$\times$6 unit cells gives pore diameters of 0.36 nm and 0.83 nm. The 12$\times$12 and 18$\times$18 configurations allow us to investigate the effect of the pore size and the pore density on the electronic properties of H-GNMs.

\begin{figure}[h]
\includegraphics[width=0.8\textwidth]{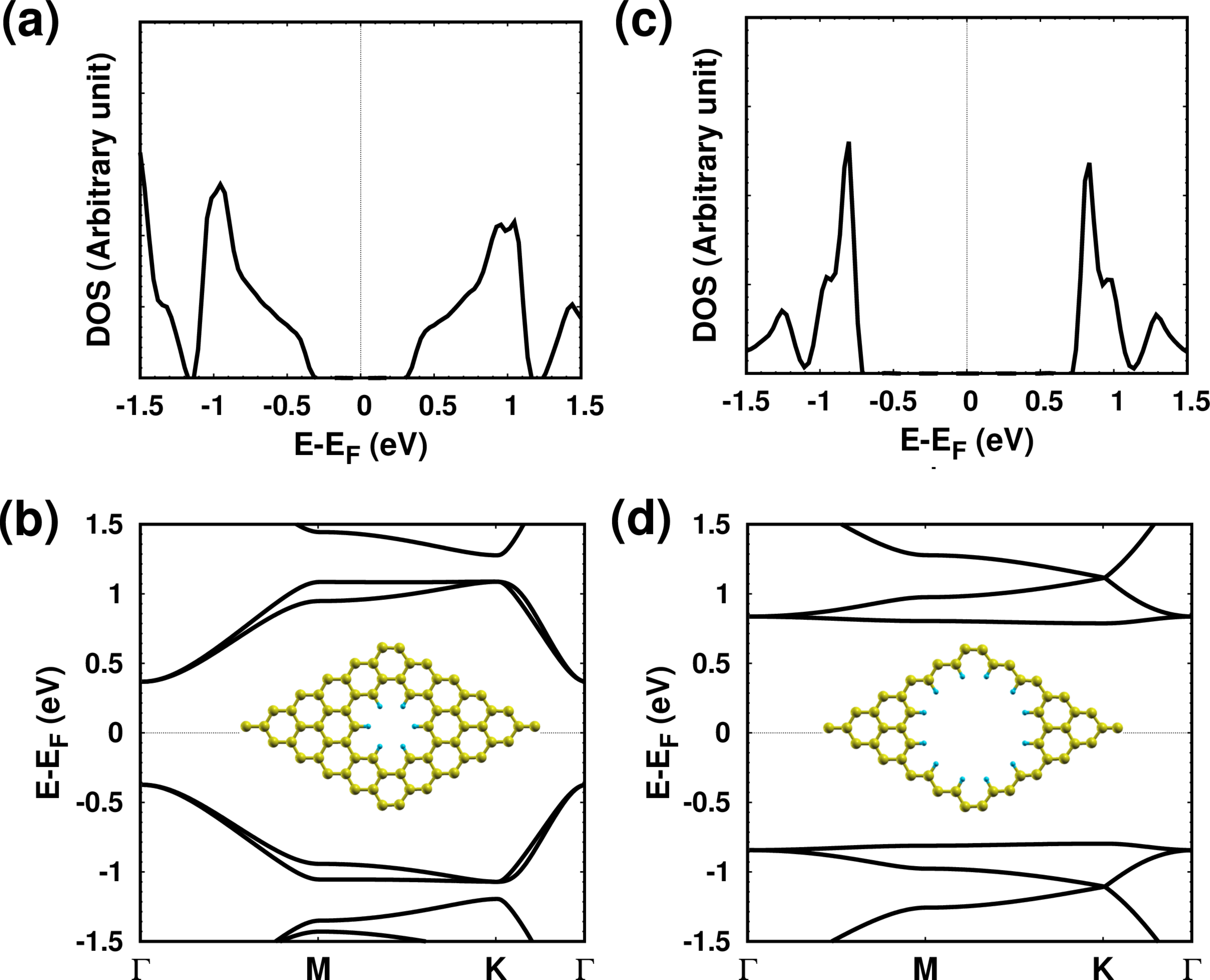}
\caption{DOS and band structure for 6$\times$6 H-GNM. (a), (b) for small pore, and (c), (d) for medium pore. The energy is given relative to the Fermi energy (E$_{F}$). The inset figures indicate the corresponding optimized structures.} 
\label{fig2}
\end{figure}

The DOS and the band structures of 6$\times$6 H-GNMs are shown in Fig. \ref{fig2}. Both H-GNMs are semiconducting with electronic bandgaps of 0.75 eV, and 1.55 eV, for small and medium pore H-GNMs, respectively. The electronic structures and bandgap values agree well with previous DFT calculations \cite{gap}. A comparison of DFTB and DFT results is discussed below. For the small pore H-GNM, the states in the vicinity of the bandgap are delocalized, and the DOS grows almost linearly away from the valence and conduction band edges. The medium pore H-GNM has two flat bands surrounding its bandgap, indicating the presence of states localized at pore edge.

\begin{figure}[h]
\includegraphics[width=0.8\textwidth]{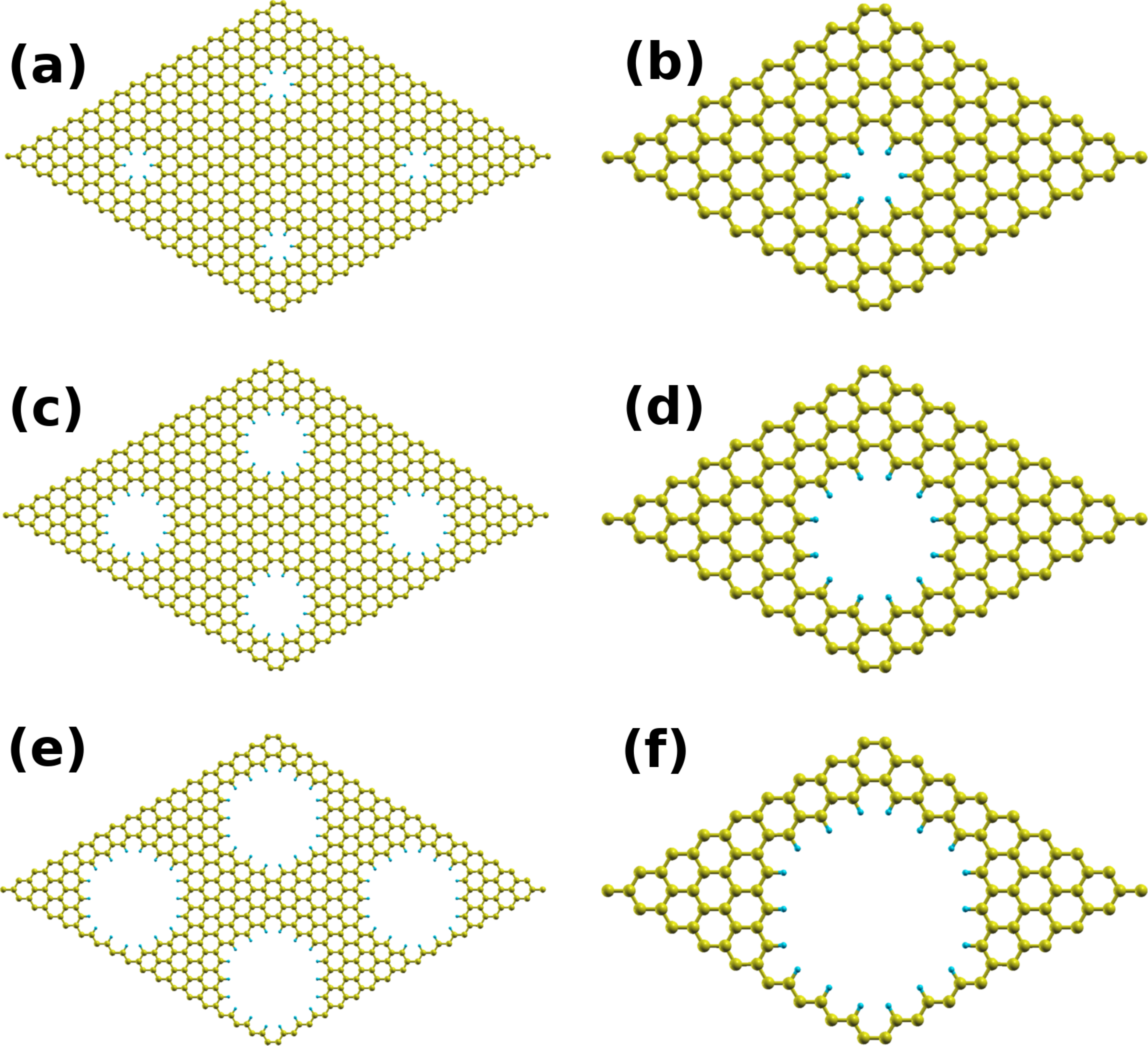}
\caption{The H-GNMs optimized structures for 18$\times$18 systems made of four 9$\times$9 unit cells, and their reduced structure, with (a), (b) small pore, (c), (d) medium pore, and (e), (f) large pore. The yellow and cyan spheres indicate C and H atoms, respectively.}
\label{fig3}
\end{figure}

We also consider cases with a different pore density (Fig. \ref{fig3}a, c, and e), where we have 18$\times$18 systems with 4 pores, and with 3 different pore sizes. When all 4 pores are equal, the unit cells reduce to the 9$\times$9 cells shown in Fig. \ref{fig3}b, d, and f. The percentages of the removed carbon atoms are 3.7\%, 14.8\%, and 33.3\%, for the small, medium, and large pore H-GNM, respectively. Structural optimization of these H-GNMs gives pore diameters of 0.36 nm, 0.83 nm, and 1.21 nm. The corresponding electronic bandgaps are 0.35 eV, 0.65 eV, and 0.82 eV, respectively. 
 
\begin{figure}[!h]
\includegraphics[width=0.8\textwidth]{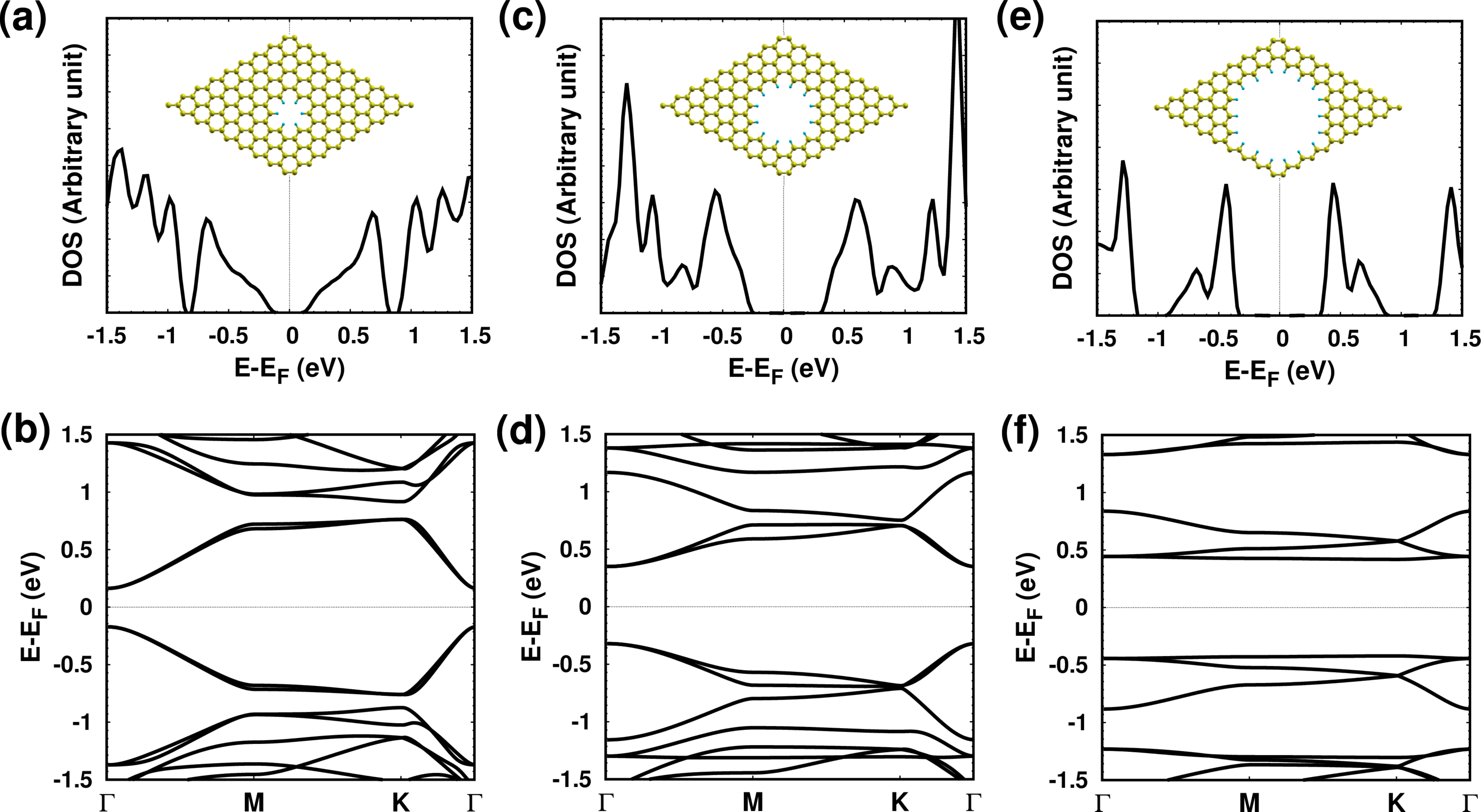}
\caption{DOS and band structure for 9$\times$9 H-GNM. (a), (b) for small pore, (c), (d) for medium pore, and (e), (f) for large pore. The energy is given relative to the Fermi energy (E$_{F}$). The inset figures indicate the corresponding optimized structures. }
\label{fig4}
\end{figure}

The DOS, the band structures (Fig. \ref{fig4}), and the bandgap values agree well with previously published DFT results \citep{AA}. The small and medium pore H-GNMs have qualitatively similar DOS and band structures, characterized by a low energy linear DOS, with delocalized states. The bandgap of the large pore H-GNM is bounded by flat bands arising from states localized at the pore edge. It should be noted here that flat bands are observed in the 6$\times$6 medium pore H-GNM, and in the 9$\times$9 large pore H-GNM, where both H-GNMs have 33.3\% of their carbon atoms removed. The presence of the localized states in these two cases can be explained by edge effects \cite{gap2}.

\begin{figure}[h]
\includegraphics[width=0.8\textwidth]{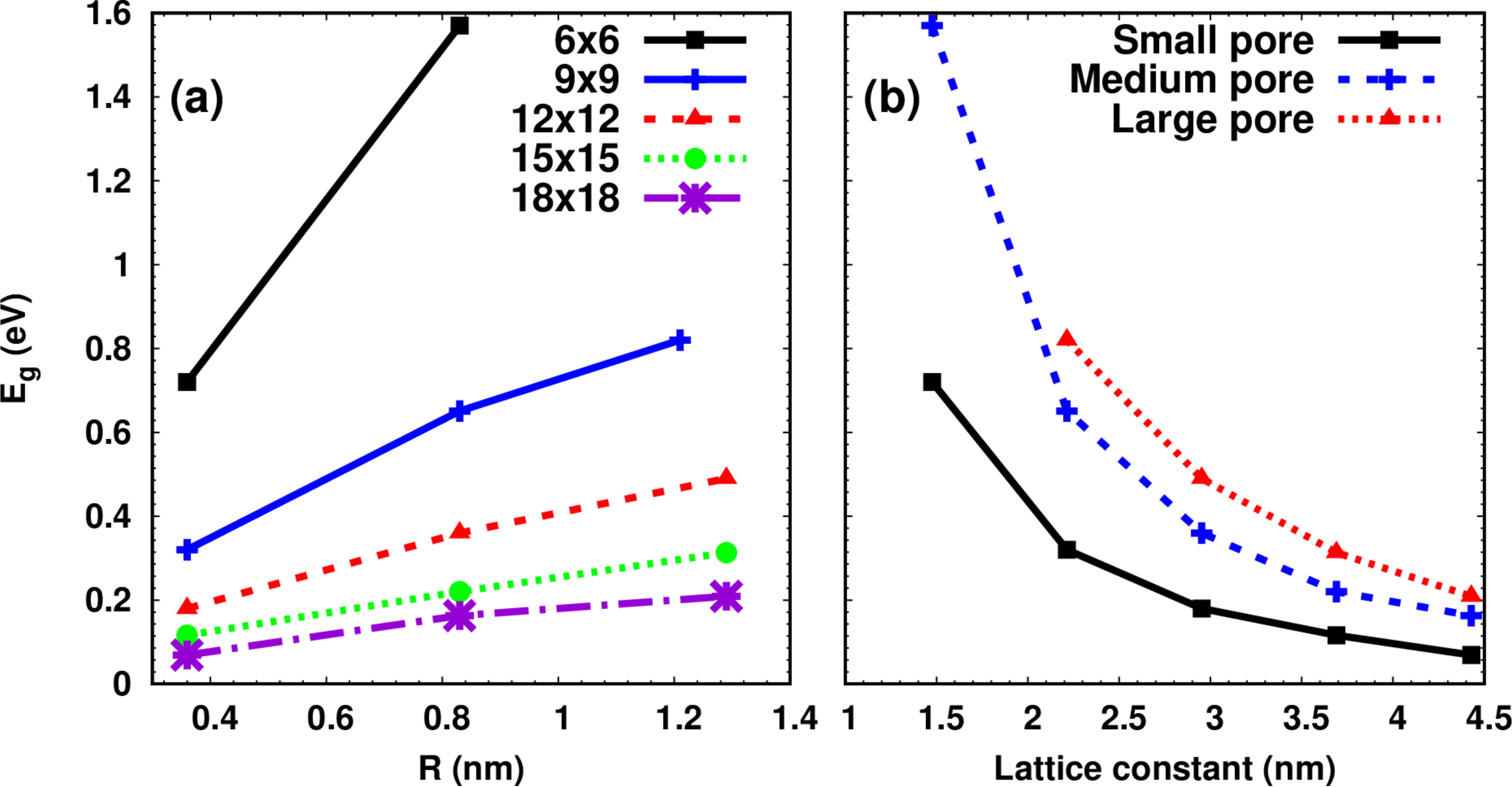}
\caption{The dependence of the bandgap on (a) the pore diameter, R, and (b) the lattice parameter.} 
\label{fig5}
\end{figure}

In addition to the systems with 6$\times$6 and 9$\times$9 unit cells discussed so far, we have also studied 12$\times$12, 15$\times$15, and 18$\times$18 systems, with each of the 3 considered pores. This allows us to further confirm the suitability of the DFTB method for studying these H-GNM systems. The dependence of the electronic bandgap on the pore diameter in each of these systems is shown in Fig. \ref{fig5}a, where we get the expected trend for the bandgap \cite{gap1,gap2}. The same results can be analyzed with respect to the pore lattice constant (Fig. \ref{fig5}b), and the bandgap can be fitted to the form: $E_g = (8.27/M^2) \exp(x)$ \cite{gap2}, where $x = 0.40 R_{a} - 0.03R_{a}^2$, $R_{a}= (2r/a_0-1)$, and $r$ is the pore radius, $a_{0}=2.64$ {\AA} is the graphene lattice constant, and $M$ indicates the periodicities of the unit cell of the GNM. It should be noted here that the 15$\times$15 and 18$\times$18 systems are almost impossible to be studied with DFT calculations due to their huge size and number of electrons.

The pore radius and the pore lattice parameter define the {\it neck width} of the GNM, which is the smallest graphene thickness between two adjacent pore edges. For a given pore radius, GNMs with certain neck widths will have their bandgaps surrounded by flat bands resulting from states localized at the pore edge \cite{gap2}. We see examples of this case in Figs. \ref{fig2}d and \ref{fig4}f.

\begin{figure}[h]
\includegraphics[width=0.9\textwidth]{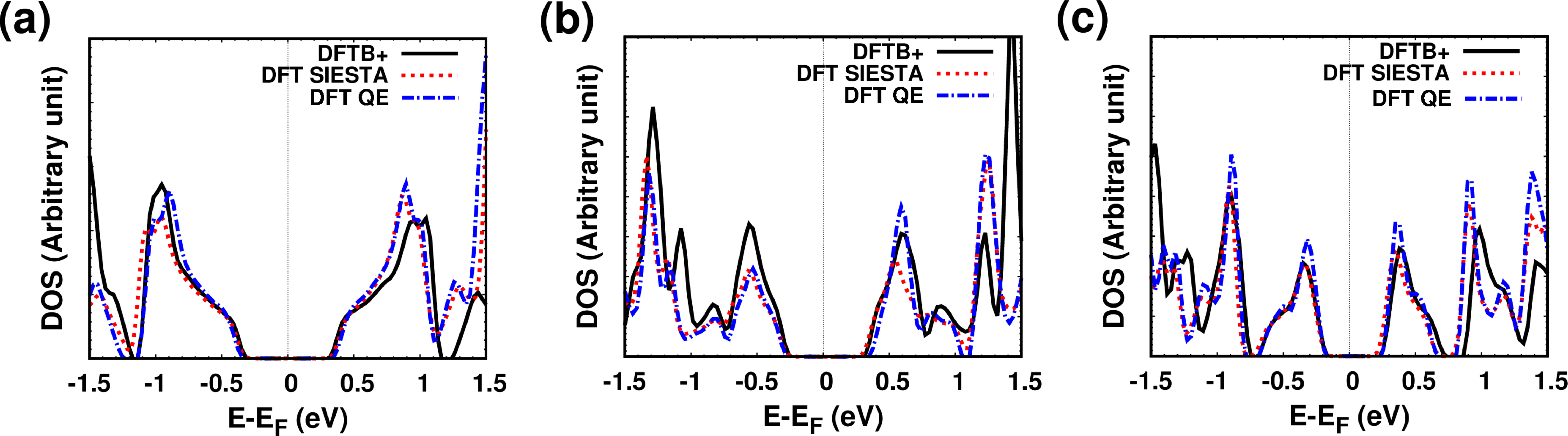}
\caption{DOS of (a) 6$\times$6 small pore, (b) 9$\times$9 medium pore, and (c) 12$\times$12 large pore systems using DFTB, and DFT.} 
\label{fig5a}
\end{figure}

Electronic properties of graphene based systems using the DFTB method have been shown to agree well with those obtained from DFT-based methods \cite{meth7}. Nevertheless, we to further validate our approach, we still compare the electronic properties of GNM basic systems used in this work, using DFTB, and DFT with two types of bases. Figure \ref{fig5a} shows the DOS of three GNMs, 6$\times$6 small pore, 9$\times$9 medium pore, and 12$\times$12 large pore, calculated using DFTB+, SIESTA (DFT, localized orbitals basis) \cite{amal}, and Quantum Espresso (DFT, plane wave basis) \cite{gap,litho1,Liang2010}. Table \ref{table_bandgap} shows the bandgap values obtained using these packages. As we see, the DFTB DOS results and the bandgap values agree well with those of DFT, which demonstrates the advantage of the DFTB method to study relatively larger systems that are formidable to DFT calculations. Although hybrid functionals usually provide more accurate bandgap estimates, the system sizes considered in our study are beyond the reach of those functionals. Nevertheless, the bandgap of the 6$\times$6 small pore system calculated using the HSE functional was found to be 0.88 eV, i.e. slightly larger than the GGA value of 0.75 eV (supplementary information of Ref.~\cite{gap}).

\begin{table}[ht]
\begin{center}
\begin{tabular}{ |c|c|c|c| }
\hline
GNM (pore size) &\multicolumn{3}{|c|}{Bandgap (eV)}\\
\hline
         & DFTB+  & SIESTA &  QE \\
\hline

$6 \times 6$ (Small)    & 0.75   &  0.70  &   0.75   \\
\hline

$9 \times 9$ (Medium)   & 0.65  &   0.60   & 0.70   \\
\hline

$12 \times 12$ (Large)  & 0.36  &  0.35   &  0.4    \\
\hline

$15 \times 15$ (Large)  & 0.31  &   0.26   & --    \\  
\hline
 
\hline
\end{tabular}
\caption{Bandgap values for 4 GNM systems obtained using DFTB and DFT.}
\label{table_bandgap}
\end{center}
\end{table}

\subsection{GNMs with different pore sizes - Structural disorder}

Now that we have established a baseline that we can compare to, we move to systems with unit cells having more than one pore size. As explained before, these systems are built from the base 6$\times$6 and 9$\times$9 systems. We begin by considering H-GNMs with small and medium pores. We have two supercell cases, 12$\times$12 (Fig. \ref{fig1}a,d), and 18$\times$18 (Fig. \ref{fig1}b,e). 

\begin{figure}[h]
\includegraphics[width=0.8\textwidth]{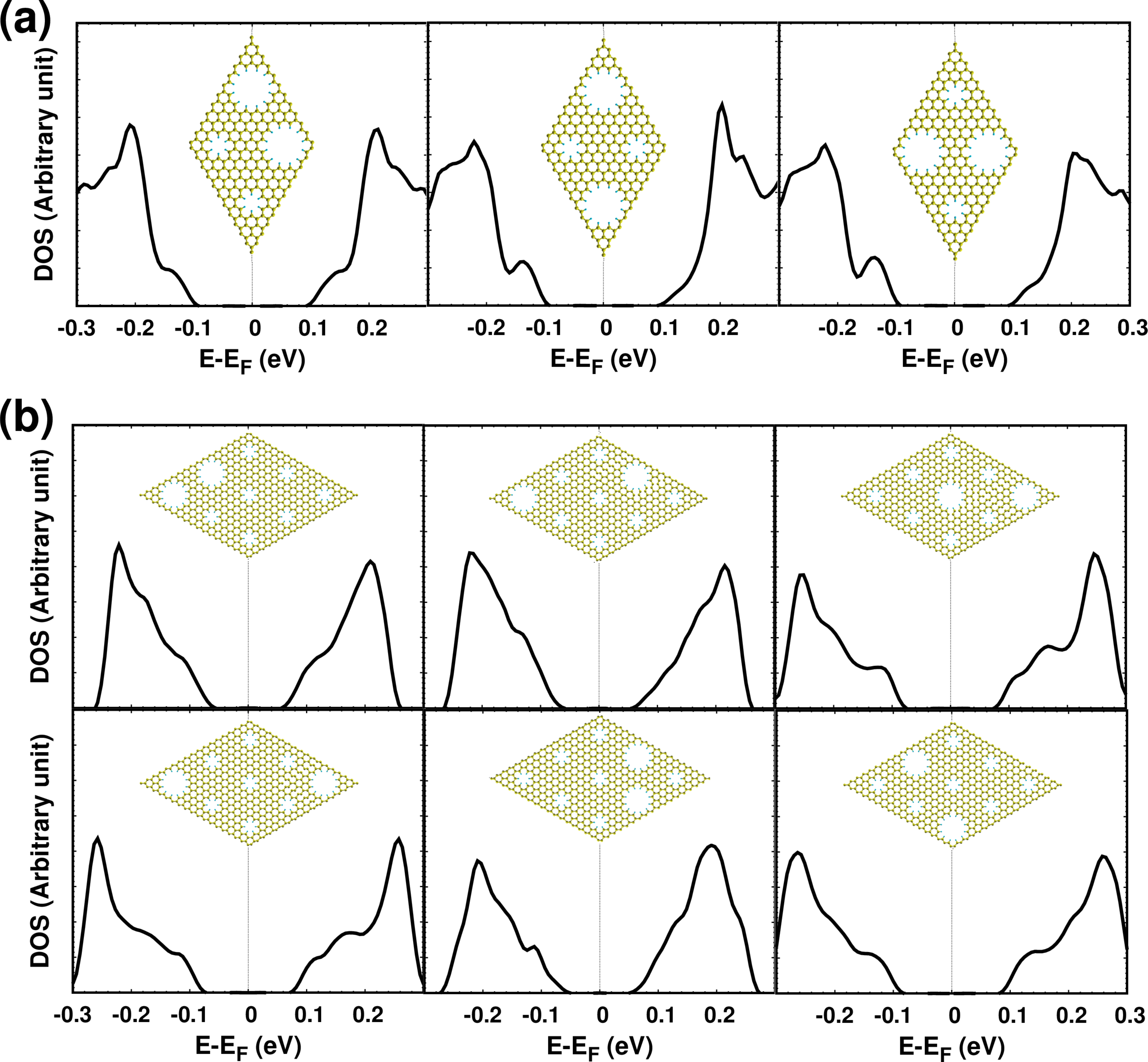}
\caption{DOS of different permutations of (a) 12$\times$12, and (b) 18$\times$18 H-GNMs, with four, and nine medium-small pores, respectively. The energy is given relative to the Fermi energy (E$_{F}$). }
\label{fig6new}
\end{figure}

In the 12$\times$12 case, we will assume we have two small pores and two medium pores. There are 3 different configurations for such a H-GNM, arising from the different permutations of the pores in the supercell (insets, Fig. \ref {fig6new}a). As we see, all three H-GNMs have virtually equal bandgaps ($E_g \sim$ 0.21 eV), and identical DOS, which suggests that the electronic properties of H-GNMs might not be highly dependent on the {\it fine} details of their structural disorder. 

To further test this result, we study the case of 18$\times$18 H-GNM with 7 small and 2 medium pores. Initially one may think that there are many permutations, but the triangular symmetry of the H-GNM supercell renders many of them identical. Figure \ref{fig6new}b shows the 6 different configurations of this H-GNM system and their DOS. The bandgap ranges between 0.12 eV and 0.18 eV for all those systems, and their electronic spectra are very similar. This further suggests that, as a first approximation, one may safely assume that the fine details of the structural disorder do not significantly affect the electronic properties of disordered H-GNMs. Therefore, we will consider only one configuration when varying the structural disorder of each H-GNM system 

We first consider H-GNMs systems with 4 pores, and with two pore sizes. Pore size can be small, medium, or large. We begin by systems where all 4 pores have one size. Different disorder configurations are obtained by continuously replacing one of the 4 pores with a pore of a different size, until we reach the other limiting case of 4 pores of the second size, with a total of 5 configurations. Each disordered system is structurally optimized, and its band structure and DOS are calculated. The disorder may not only affect the size of the electronic bandgap, and consequently the ON/OFF ratio of an H-GNM-based field effect transistor (FET) device \cite{synth3,transistor}, but it can also lead to a degradation of the electronic mobility through a diminished carrier group velocity \cite{gap}. To investigate this effect, we calculated the group velocity of carriers in the linear region of the conduction band of various studied H-GNMs.

\begin{figure}[h]
\includegraphics[width=0.8\textwidth]{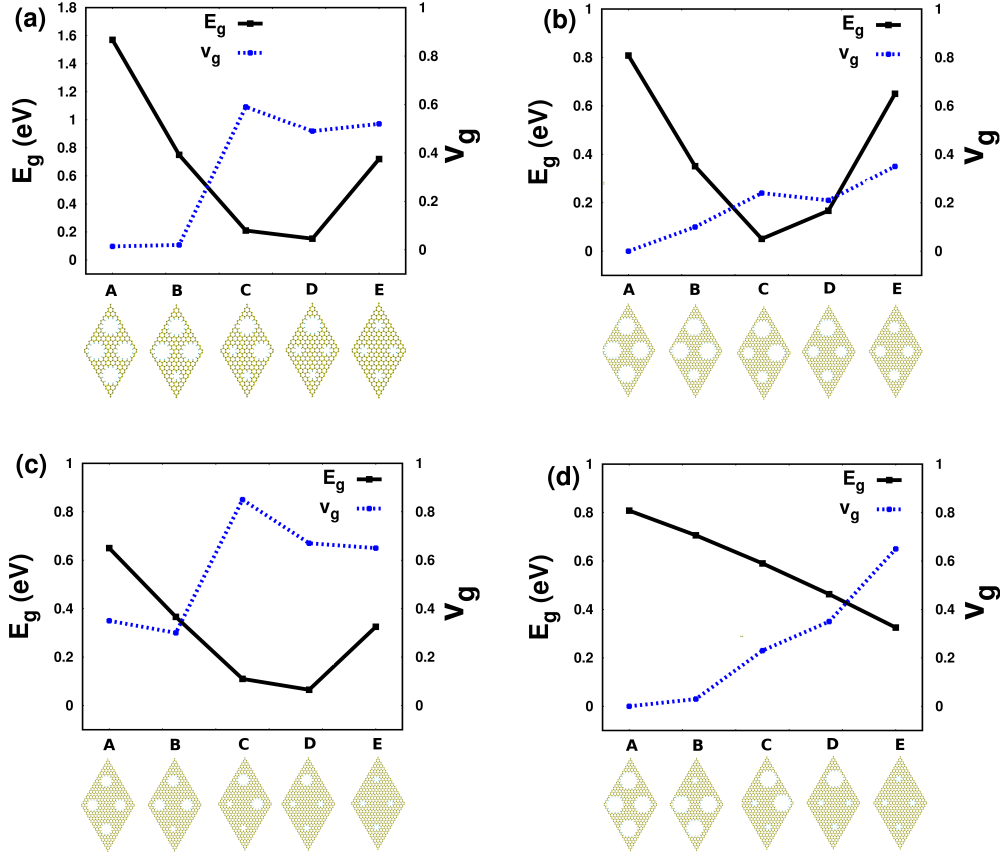}
\caption{Bandgap and relative group velocity of different disordered structures for (a) 12$\times$12 H-GNM with medium-small pores, 18$\times$18 H-GNM with (b) large-medium pores, (c) medium-small pores, and (d) large-small pores.}
\label{fig7}
\end{figure}

Figure \ref{fig7} shows the variation of $E_g$ (left $y$-axis) and $v_g$ (right $y$-axis) of the 12$\times$12 and 18$\times$18 H-GNMs with different disorder configurations. We design the plots such that the two disorder free structures are the limiting cases on the far left and far right of the $x$-axis. Figure \ref{fig7}a shows $E_g$ and $v_g$ of the structurally disordered 12$\times$12 H-GNM system, with small and medium pores. The bandgap decreases between the two disorder-free values of 1.55 eV and 0.72 eV, with the smallest bandgap of 0.15 eV for the H-GNM system with 3 small pores and one medium pore (structure D). The group velocity follows a trend that is roughly proportional to the bandgap, and ranges between 0.2 and 0.8 of the velocity of pristine graphene, which is still attractive from a FET device perspective.

Figure \ref{fig7}b shows $E_g$ and $v_g$ of the disordered 18$\times$18 H-GNM system made from medium and large pores. Structural disorder causes the bandgap to decrease below the two disorder-free limiting cases (0.8 eV and 0.65 eV), with the smallest bandgap (0.07 eV) being for the H-GNM with one medium and three small pores (structure C). For the two other disordered systems (B and D), $E_g \sim 0.2$ eV and $v_g \sim 0.1$. Figure \ref{fig7}c shows that the bandgap of the 18$\times$18 H-GNM with medium and small pores follows a similar trend, where it decreases between the two limiting cases of 0.65 eV and 0.34 eV. The maximum decrease is for the H-GNM system having one medium and three small pores, with $E_g \sim 0.1$ eV, and a $v_g \sim 0.1$ (structure D). The last 18$\times$18 4 pore H-GNM system is that with large and small pores (Fig. \ref{fig7}d). $E_g$ maintains a monotonic decrease between the disorder-free values of 0.8 eV and 0.34 eV, and $v_g \geq 0.2$ for two of the three disorder cases.  

\begin{figure}[h]
\includegraphics[width=0.8\textwidth]{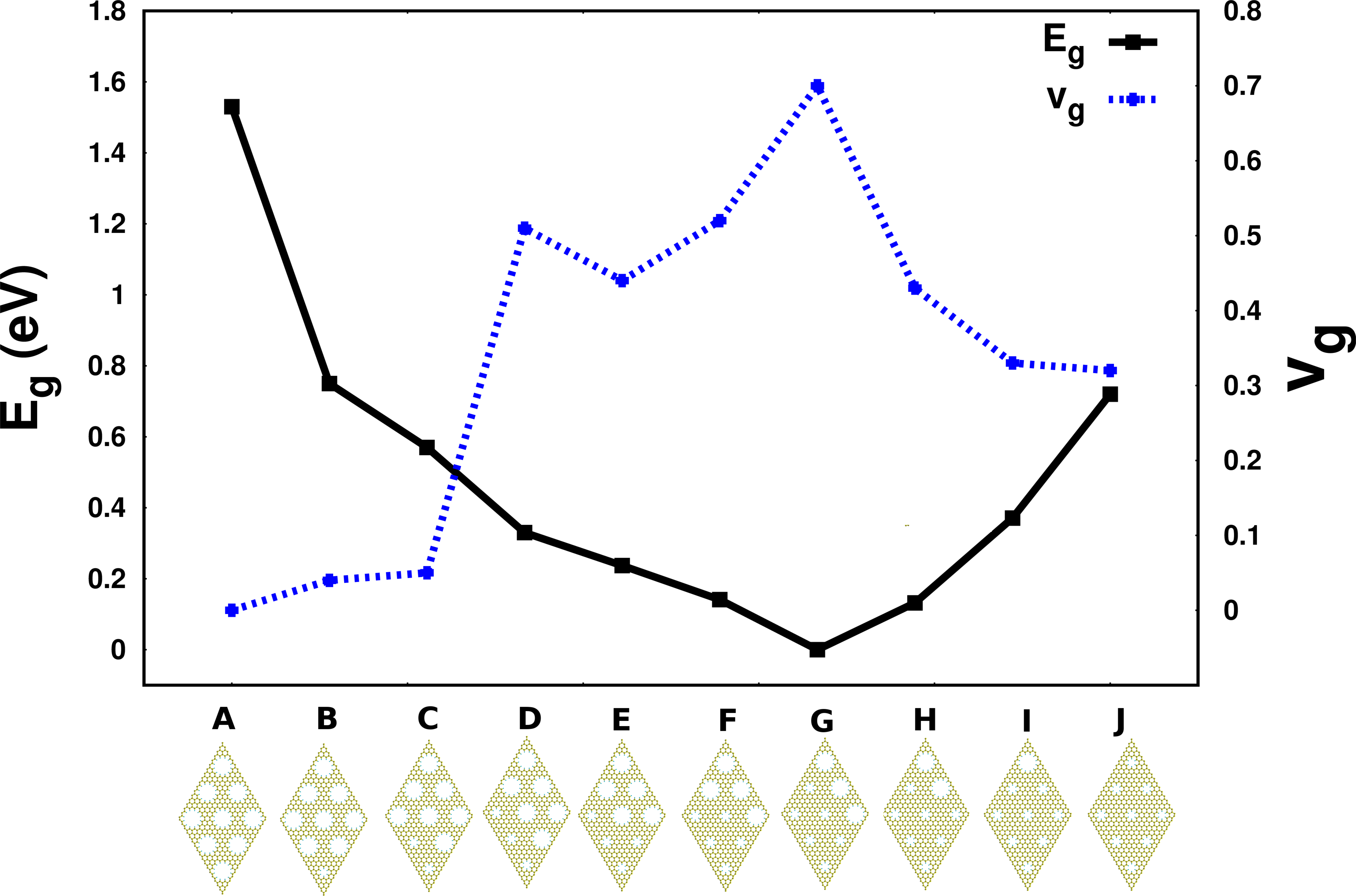}
\caption{Bandgap and relative group velocity of different disordered structures for 18$\times$18 H-GNM with medium-small pores.}
\label{fig8new}
\end{figure}

We now move to the 18$\times$18 H-GNM systems with 9 pores, and two pore sizes, small and medium. We introduce disorder by following a procedure similar to that of the 4-pore H-GNM systems. The larger number of pores allows for a larger number of disordered H-GNM configurations. The disorder axis (Fig. \ref{fig8new}, $x$-axis) has the H-GNM with 9 medium pore on the far left (point A), and the H-GNM with the 9 small pores on the far right (point J), with 8 disordered configurations in between. These configurations are obtained by continuously replacing one medium pore with one small pore. The $y$-axis is as before. $E_g$ exhibits a trend similar to that of the 12$\times$12 case, decreasing between the two disorder free values of 1.55 eV and 0.72 eV, and vanishing for the structure with 3 medium and 6 small pores (point G). For all other disordered cases, $E_g$ is between 0.2 and 0.8 eV, and $v_g$ is between 0.2 and 0.7, both are very lucrative from a device perspective.

Our calculations may help interpret the recent experimental transport measurements made on GNMs by Schmidt et al \cite{Schmidt2018}. They used helium ion beam milling to pattern suspended graphene, creating pores with diameters of about 3.5 nm, and with a pitch of about 18 nm. They measured the I/V characteristics of suspended GNMs at room temperature, reporting a gap of about 0.45 eV for one device. Their results include another GNM device with the same pore parameters, but with some {\it missing} pores, was found to have no gap. This agrees well with our theoretical predictions, and provides experimental evidence that disorder in pore lattice may cause a reduction in the bandgap of a GNM. 

\begin{figure}[h]  
\includegraphics[width=0.8\textwidth]{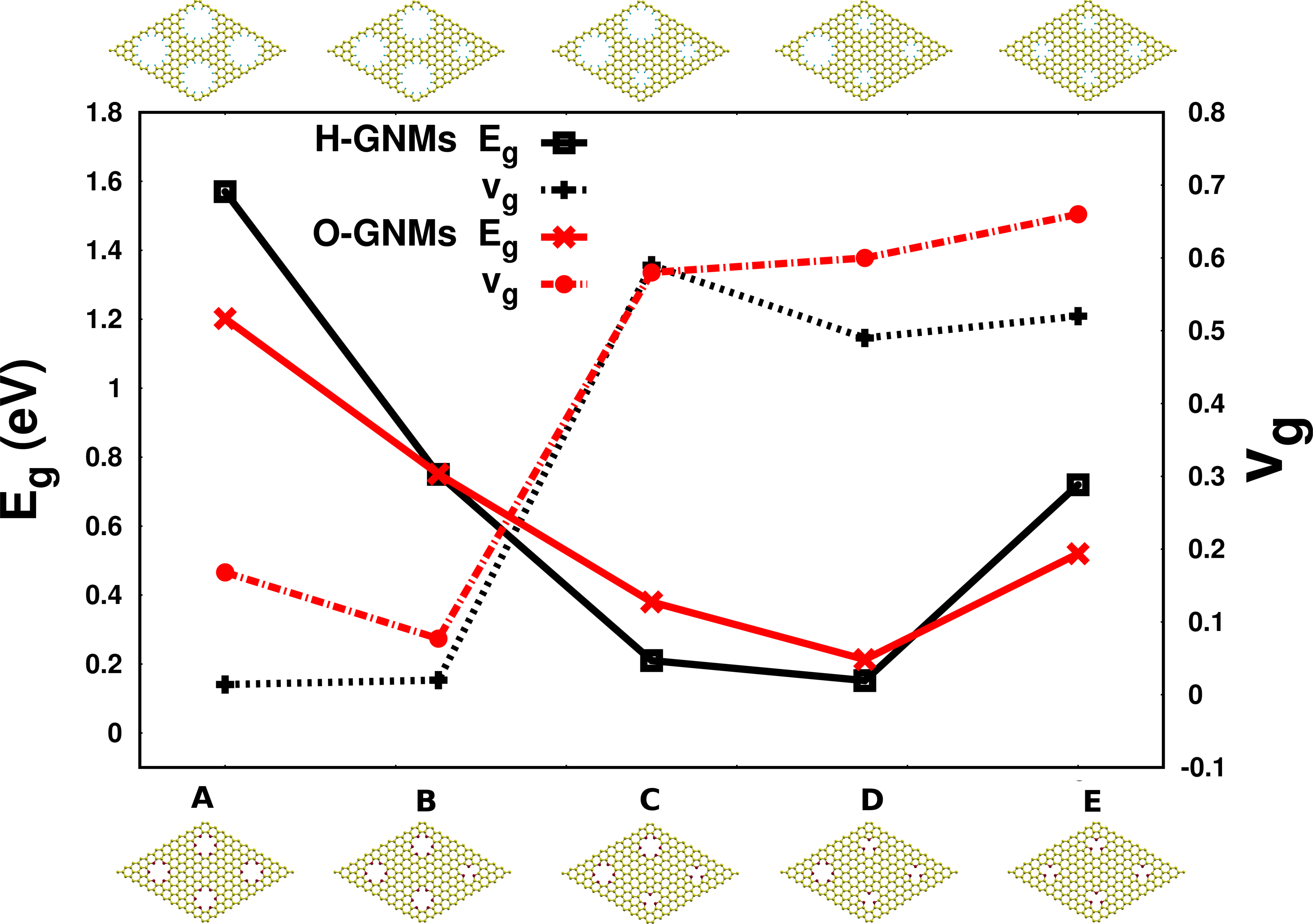}
\caption{Bandgap and relative group velocity of different disordered structures for 12$\times$12 for H-GNM (upper axis) and O-GNM (lower axis), with medium-small pores (see table \ref{t1}). The yellow, cyan, and red spheres in the optimized structures indicate C, H, and O atoms, respectively.}
\label{fig9} 
\end{figure}

All systems presented so far had their pore edges passivated with hydrogen. In general, and irrespective of how the GNM is fabricated, the pores will be passivated by some chemical species. The most common passivations considered are hydrogen, oxygen, and nitrogen \cite{gap,hex1}. For this, we studied a set of oxygen-passivated GNMs. Figure \ref{fig9} compares the dependence of the electronic bandgap on structural disorder for an oxygen-passivated and the hydrogen passivated 12$\times$12 GNM systems. As we see, the trend is not altered by the change in passivation, and the variations in the values of the bandgap can be attributed to the local deformations occurring near the passivation sites. The group velocity is generally between 0.2 and 0.9 of that in graphene, which is very pleasing from a device perspective. In certain cases though flat bands exist due to states localized at the pore edge, in which case the group velocity is very small. 

Other forms of disorder that may be considered in future work include pores with unequal number of removed $A$ and $B$ sublattices carbon atoms. This may arise in many situations, such as with irregularly shaped pores  \cite{gap2, Sun2019}, triangular pores \cite{PhysRevB.84.214404}, or with circular/hexagonal pores centered on carbon atoms. This results in the presence of midgap localized states, as well as magnetism with a magnetic moment that is proportional to the difference between the $A$ and $B$ atoms. This is not unique to graphene, but rather to bipartite lattices, as predicted by Lieb's theorem \cite{Lieb1989}.
 
\section{Conclusion}
 
The electronic properties of structurally disordered graphene nanomeshes (GNMs) were studied using the density functional based tight binding method. We defined structurally disordered GNM as one whose unit cell has two different pore sizes. We calculated band structures and densities of states for 56 GNM systems, with different pore sizes, different permutations of the pores, and different passivations (hydrogen and oxygen). We found that the disorder in the pore size causes the energy gap to decrease, albeit to a value that is still technologically attractive. In addition, the carriers group velocity is typically between 0.2 and 0.9 of the velocity in pristine graphene, again suggesting the plausibility of a GNM-based field effect transistor. Furthermore, changing the location of pores in the unit cell seems to have little effect on the electronic properties of a GNM. These results apply for the two considered pore passivations. Our results indicate that the utilization of GNMs for nanoelectronic devices is possible, even with some pore size variability resulting during the fabrication process.
 
\section*{Acknowledgment}
The authors would like to acknowledge the support of the
supercomputing facility at the Bibliotheca Alexandrina, Alexandria,
Egypt. We would also like to thank B. Aradi and Ben Hourahine for fruitful
discussions. 
 
\bibliography{bibfile1} 
\bibliographystyle{elsarticle-num-names}
	
\end{document}